\documentclass{aa}

\usepackage{graphicx}
\usepackage{amssymb}
\usepackage{latexsym}
\usepackage{latexsym}
\usepackage{graphics}
\usepackage{txfonts}

\makeatletter
\renewcommand\@biblabel[1]{}
\makeatother

\def\ha{H$\alpha$}
\def\hb{H$\beta$}

\def\niib{[NII]$\lambda$6584}
\def\niia{[NII]$\lambda$6548}

\def\oiiia{[OIII]$\lambda$4959}
\def\oiiib{[OIII]$\lambda$5007}

\def\siia{[SII]$\lambda$6716}
\def\siib{[SII]$\lambda$6730}
\def\siiab{[SII]$\lambda$6716,30}
\def\oi{[OI]$\lambda$6300}

\def\heib{[HeI]$\lambda$6678}
\def\heic{[HeI]$\lambda$7065}
\def\siii{[SIII]$\lambda$9069}

\newcommand {\mean} [1] {\langle#1\rangle}

\begin{document}


\title{The MUSE view of He~2-10: 
no AGN ionization \\but a sparkling starburst
		\thanks{This work is based on observations made at the European Southern Observatory, Paranal, Chile (ESO program 095.B-0321)}
}
	
\author{G. Cresci\inst{1}\fnmsep, 
L. Vanzi\inst{2},
E. Telles\inst{3},
G. Lanzuisi\inst{4,5},
M. Brusa\inst{4,5},
M. Mingozzi\inst{5,1},
M. Sauvage\inst{6},
K. Johnson\inst{7}
}

\institute{INAF - Osservatorio Astrofisco di Arcetri, largo E. Fermi 5, 50127 Firenze, Italy \\ \email{gcresci@arcetri.astro.it}  
	\and
	Department of Electrical Engineering and Center of Astro Engineering, Pontificia Universidad Catolica de Chile, Av. Vicu\~na Mackenna 4860, Santiago, Chile
	\and
	Observatorio Nacional, Rua Jos\'e Cristino 77, 20921-400 Rio de Janeiro, Brasil 
	\and
	Dipartimento di Fisica e Astronomia, Universit\`a di Bologna, viale Berti Pichat 6/2, 40127 Bologna, Italy
	\and
	INAF - Osservatorio Astronomico di Bologna, via  Gobetti 93/2, 40129 Bologna, Italy
	\and
	Laboratoire AIM, CEA/IRFU/Service d’Astrophysique, CNRS, Universit\'e Paris Diderot, Bat. 709, 91191 Gif-sur-Yvette, France 
	\and
	Department of Astronomy, University of Virginia, 530 McCormick Road, Charlottesville, VA 22904, USA	\\
}

\date{Received ; accepted }

\abstract{
We study the physical and dynamical properties of the ionized gas in the prototypical HII galaxy Henize~2-10 using MUSE integral field spectroscopy. The large scale dynamics is dominated by extended outflowing bubbles, probably the results of massive gas ejection from the central star forming regions. We derive a mass outflow rate $\dot M_{out}\sim 0.30\ M_{\sun}\ yr^{-1}$, corresponding to mass loading factor $\eta\sim0.4$, in range with similar measurements in local LIRGs. Such a massive outflow has a total kinetic energy that is sustainable by the stellar winds and Supernova Remnants expected in the galaxy. We use classical emission line diagnostic to study the dust extinction, electron density and ionization conditions all across the galaxy, confirming the extreme nature of the highly star forming knots in the core of the galaxy, which show high density and high ionization parameter. We measure the gas phase metallicity in the galaxy taking into account the strong variation of the ionization parameter, finding that the external parts of the galaxy have abundances as low as  $12 + log(O/H) \sim 8.3$, while the central star forming knots are highly enriched with super solar metallicity.
We find no sign of AGN ionization in the galaxy, despite the recent claim of the presence of a super massive active Black Hole in the core of He~2-10. We therefore reanalyze the X-ray data that were used to propose the presence of the AGN, but we conclude that the observed X-ray emission can be better explained with sources of a different nature, such as a Supernova Remnant.
}


\authorrunning{Cresci et al.}

\maketitle

\section{Introduction}

Nearby HII galaxies represent an unique test bench to study in detail the physical mechanisms that drive star formation and galaxy evolution in nearly pristine environments, resembling those in high-z galaxies. In fact, their low chemical abundances, high gas fractions, high specific star formation rate (sSFR) and the dominance of very young stellar populations in massive star clusters allow a closer comparison with the conditions present in primordial star forming galaxies at high redshift. Given their proximity, these processes can be studied at much higher spatial resolution than at high-z, thus representing a fundamental test in our understanding of galaxies and of galaxy evolution. For these reasons our group has undertaken a long term program to investigate the internal physical properties of these star forming galaxies through spatially resolved spectroscopy in the optical and in the near-IR, in order to assess their kinematics, physical conditions and their relation to their local star formation properties (see e.g. Vanzi et al.~\citealp{vanzi08}, \citealp{vanzi11}; Cresci et al. \citealp{cresci10}; Lagos et al. \citealp{lagos12}; Telles et al. \citealp{telles14}).

With the advent of the new large field Multi Unit Spectroscopic Explorer (MUSE, Bacon et al. \citealp{bacon10}) optical integral field spectrometer, we target Henize~2-10, one of the prototypical HII galaxies (Allen et al. \citealp{allen76}). It is located at just 8.23 Mpc (Tully et al. \citealp{tully13}, providing a scale of $40$ pc arcsec$^{-1}$), and its optical extent is less than 1 kpc, showing a complex and irregular morphology. Despite its low mass ($M_*=3.7\times 10^9\ M_{\sun}$, Reines et al. \citealp{reines11}, although Nguyen et al. \citealp{nguyen14} revised this value to $10\pm 3\times10^9\ M_{\sun}$ using deep optical imaging of the outer regions), it hosts an intense burst of star formation (Star Formation Rate SFR$=1.9\ M_{\sun}\ yr^{-1}$, Reines et al. \citealp{reines11}), as traced by classical indicators such as strong optical and infrared emission lines excited by young stars (Vacca \& Conti \citealp{vacca92}, Vanzi \& Rieke \citealp{vanzi97}), the detection of Wolf-Rayet features (Schaerer et al. \citealp{schaerer99}), intense mid-IR emission (Sauvage et al. \citealp{sauvage97}, Vacca et al. \citealp{vacca02}) and far-IR continuum (Johansson \citealp{johansson87}). 
The galaxy is gas rich, with a molecular gas mass of $\sim 1.6\times 10^8\ M_{\sun}$  and an atomic gas mass of $1.9\times 10^8\ M_{\sun}$ (Kobulnicky et al. \citealp{kobulnicky95}, suggesting a gas fraction of about 3-10\%). 

\begin{figure}
	\begin{center}
		\includegraphics[width=0.47\textwidth]{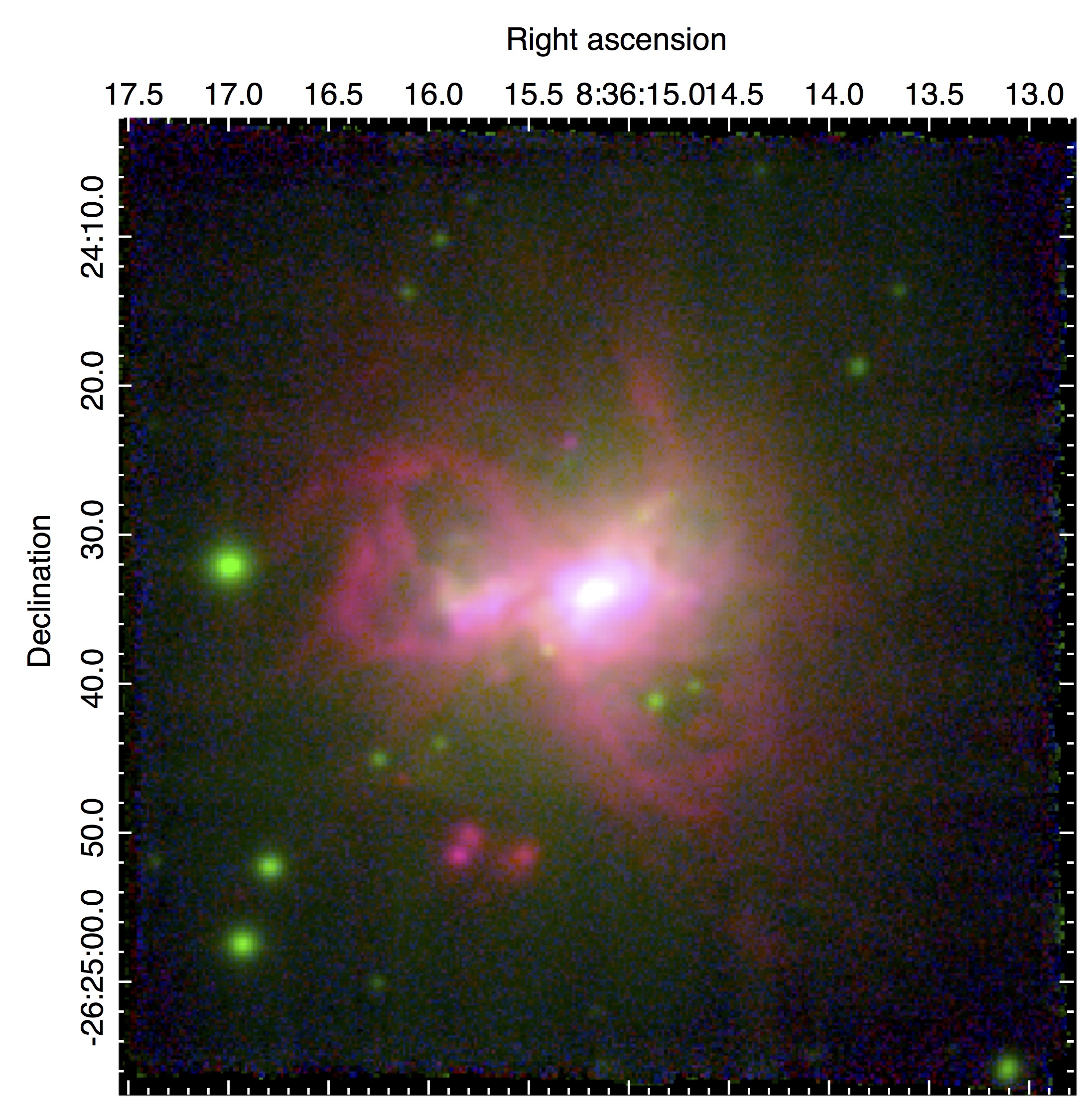}
		\includegraphics[width=0.47\textwidth]{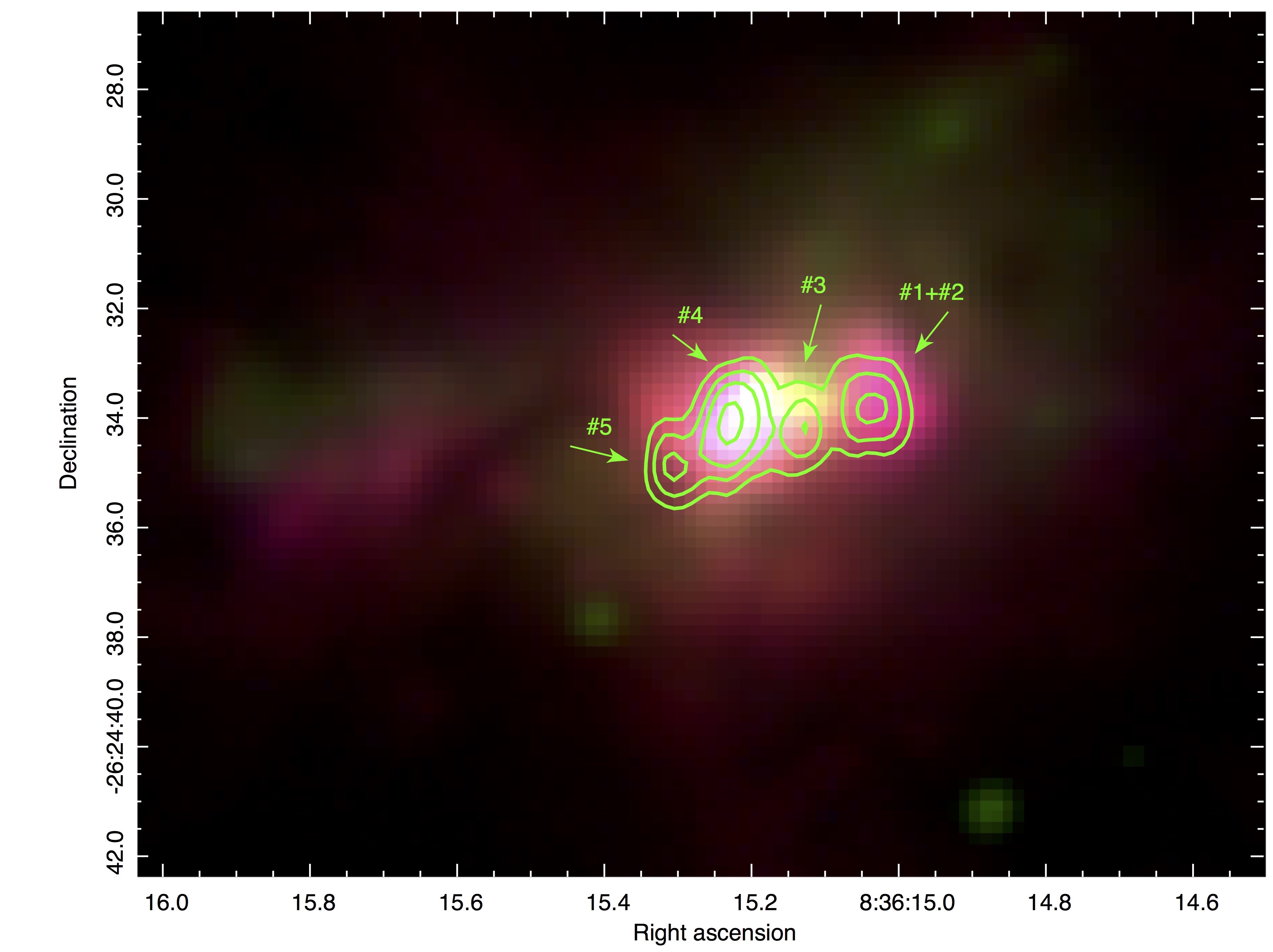}
	\end{center}
	\caption{Three color MUSE images of the galaxy He~2-10. The full MUSE field of view ($1'\times1'$) is shown in the \textit{upper panel}: the continuum over the entire spectral range is shown in green, the \oiiib\ emission in blue, and the \ha\ emission in red. An enlarged view of the central regions of the galaxy ($20''\times15''$) is instead shown in the \textit{lower panel}, with the radio continuum contours at $3.6\ cm$ obtained by Johnson \& Kobulnicky \cite{johnson03} overplotted, along with their notation for the main radio knots 1-5. North is up and East is left.} 
	\label{3colors}
\end{figure}

Apart from the central $500$ pc characterized by blue colors and irregular morphology, the outer part of the galaxy appears to be redder and dominated by an older stellar population, consistent with an early-type system (Nguyen et al. \citealp{nguyen14}). The stellar component seems to be dispersion dominated also in the central part, further supporting the spheroidal nature of the older stellar population (Marquart et al. \citealp{marquart07}). The recent burst ($\sim 10^7\ yr$, Beck et al. \citealp{beck97}) was probably triggered by a massive gas infall due to an interaction, or to a merger with a companion dwarf, as inferred by the tidal CO and HI plumes extended up to $30"$ to the southeast and northeast of the core (Kobulnicky et al. \citealp{kobulnicky95}, Vanzi et al. \citealp{vanzi09}).

The galaxy  is also surrounded by a complex kiloparsec-scale superbubble centered on the most intense star forming core, and evident in HST narrowband \ha\ images (Johnson et al. \citealp{johnson00}). The estimate of the energy of the bubbles is compatible with the expected mechanical energy released by Supernovae (SNe) and stellar winds in the central starburst, suggesting a stellar driven galaxy wind (M{\'e}ndez et al. \citealp{mendez99}). The outflowing material was also detected in absorption as blueshifted interstellar lines up to $-360\ km/s$, that were used to compute a lower limit on the mass of the outflowing material of $M_{out}\gtrsim10^6\ M_{\sun}$ (Johnson et al. \citealp{johnson00}).

The most actively star forming regions are concentrated in the core of the galaxy, in an area of about $3"$ in radius (120 pc). Here
optical and UV studies have detected an arc-like structure of resolved young super massive star clusters, with  masses up to $10^5\ M_{\sun}$ and ages of $\sim10$ Myr (e.g. Conti \& Vacca \citealp{conti94}, Johnson et al. \citealp{johnson00}, Cresci et al. \citealp{cresci10}), prompting for the first time the question of whether this is a dominant mode of star formation in galaxies. These clusters lie at the center of a cavity depleted of both cold molecular gas and warm line emitting ionized gas, whose emission is instead concentrated in two regions East and West of the arc (knot 4 and 1+2 in fig.~\ref{3colors}), harboring actively star forming, dust embedded young star clusters, prominent at mid-IR wavelength (ages $\lesssim 5\times 10^6\ yr$ and masses $\gtrsim 5\times 10^5\ M_{\sun}$, Vacca et al. \citealp{vacca02}, Cabanac et al. \citealp{cabanac05}). The peculiar configuration of a cavity with older clusters surrounded by gas rich regions was interpreted by Cresci et al. \citealp{cresci10} as an example of stellar `positive feedback', as the estimates of the shock shell velocity agree with the measured ages of the young clusters in the IR sources. All these IR clusters are associated with extended radio sources, and their properties are mostly explained as Ultra-Dense HII regions (UDHII) powering the radio emission (Kobulnicky \& Johnson \citealp{kobulnicky99}, Johnson \& Kobulnicky \citealp{johnson03}).

The only notable exception is a non-thermal radio source located in the line emitting bridge between the two \ha\ brightest clumps (designated as knot 3 in Johnson \& Kobulnicky \citealp{johnson03}, see Fig~\ref{3colors}. Note that Cabanac et al. \citealp{cabanac05} detected in L band two infrared source at this location). This source is very compact, with a physical scale of $\leq 3\ pc\ \times\ 1\ pc$ (Reines \& Deller \citealp{reines12}), and has been associated by Reines et al. \cite{reines11} to a compact X-ray emission detected with Chandra observations by Kobulnicky \& Martin \cite{kobulnicky10}. Using their combination of radio and X-ray fluxes, Reines et al. \cite{reines11} concluded that the most likely explanation for the radio source was an actively accreting super massive Black Hole (SMBH), with $log(M/M_{\sun})=6.3\pm1.1$. As the presence of very few SMBHs have been previously inferred in low mass highly star forming galaxies (see e.g. Barth et al. \citealp{barth04}), and given that He~2-10 shows no sign of a bulge or massive cluster related to the putative SMBH position, this claim has important implications on our picture of SMBH and galaxy assembly and coevolution, suggesting that black hole seeds may evolve faster than their hosts. 
However, recent and deeper follow-up Chandra observations presented in Reines et al. \cite{reines16} have shown that the hard X-ray emission previously identified was dominated by an additional source that is distinct from the compact radio source and compatible with an off-center X-ray binary. If the interpretation of the compact radio source as a SMBH is maintained, their estimate of its hard X-ray radiation without contamination brings down the accretion of the candidate SMBH well below the Eddington limit ($\sim10^{-6}\ L_{Edd}$).

Here we present new MUSE optical integral field observations of He~2-10, to study for the first time the global ionization, dynamics and physical properties of the warm ionized gas on large scales. The paper is organized as follows: in Sect. \ref{observations} we describe the MUSE observations, data reduction, and the method used for the spectral fitting of the continuum and of the emission lines; in Sect. \ref{dynamics} we discuss the obtained gas dynamics, the ionization and metallicity properties are discussed in Sect. \ref{gasmet}, and in Sect. \ref{BH} we discuss the evidences of the presence of a SMBH in He~2-10 from our MUSE data as well as archive Chandra X-ray ones. Our conclusions follow in Sect. \ref{conclusions}.

\begin{figure*}
	\begin{center}
		\includegraphics[width=0.9\textwidth]{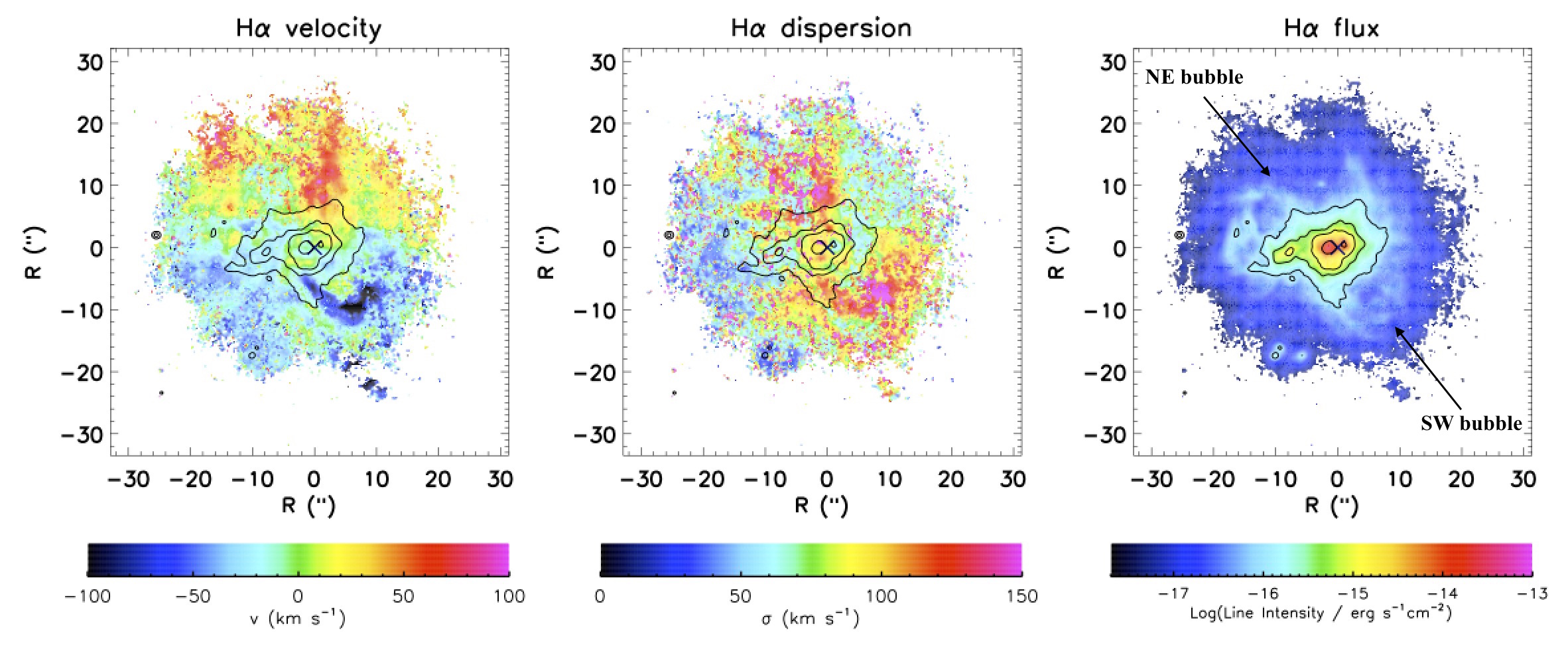} 
	\end{center}
	\caption{Gas kinematics from the combined fit of the emission lines in He~2-10. \textit{Left panel}: ionized gas velocity, relative to the systemic velocity of $873\ km\ s^{-1}$ derived from the integrated spectrum. \textit{Left panel}: $\sigma$ velocity dispersion of the total line profile, corrected for instrumental effects. \textit{Right panel}: \ha\ map, plotted as reference, with the location of the blueshifted SW bubble and redshifted NE bubble marked. The contours of the \ha\ line emission are shown in each panel, as well as the location of the radio source identified with a cross, and all maps are displaying the spaxels with $S/N>3$ in \ha. The expanding bubbles both in approaching and receding velocities are evident in the maps.} 
	\label{Hadyn}
\end{figure*}

\section{Observations and data reduction} \label{observations}

He~2-10 was observed with MUSE on May 28, 2015, under program 095.B-0321 (PI Vanzi).  The galaxy was observed with four dithered pointings of $30\ s$ each, for a total of 2 minutes on source, with the background sky sampled with 3 equal exposures in between.
The data reduction was performed using the recipes from the  latest version of the MUSE pipeline (1.6.2), as well as a collection of custom IDL codes developed to improve the sky subtraction, response curve and flux calibration of the data. Further details on the data reduction can be found in Cresci et al. \cite{cresci15b} and references therein. 
The final datacube consists of $321\times328$ spaxels, for a total of over 100000 spectra with a spatial sampling of $0.2"\times0.2"$ and a spectral resolution going from 1750 at $465\ nm$ to 3750 at $930\ nm$. The field of view is $\sim 1'\times1'$, sufficient to sample the optical extent of the source (see Fig. \ref{3colors}). The average seeing during the observations, derived directly from foreground stars in the final datacube, was $FWHM=0.68"\pm0.02"$.

\subsection{Emission line fitting} \label{lines}

The obtained datacube was analyzed using a set of custom python scripts, developed to subtract the stellar continuum and fit the emission lines with multiple Gaussian components where needed. The details about the procedures used are described in Venturi et al. (in prep.); here we just summarize the steps followed to obtain the emission lines fluxes, velocities and velocity dispersions. 

The underlying stellar continuum was subtracted using a combination of MILES templates (S{\'a}nchez-Bl{\'a}zquez et al. \citealp{SB06}) in the wavelength range $3525-7500\ \AA$ covered by the stellar library. This wavelength interval covers most of the emission lines considered in the following, with the exception of \siii, for which we used a polynomial fit to the local continuum given that this line is not contaminated by underlying absorptions. The continuum fit was performed using the pPXF code (Cappellari \& Emsellem \citealp{cappellari04}) on binned spaxels using a Voronoi tessellation (Cappellari \& Copin \citealp{cappellari03}) to achieve a minimum $S/N>50$ on the continuum under $5530\ \AA$ rest frame. The main gas emission lines included in the selected range (\hb, \oiiia, \oiiib, \heib, \oi, \niia, \ha, \niib, \heic, \siiab) were fitted simultaneously to the stellar continuum using multiple Gaussian components to better constrain e.g. the absorption underlying the Balmer lines. Fainter lines as well as regions affected by sky residuals were masked out of the fitting region. The fitted stellar continuum emission in each bin is then subtracted on spaxel to spaxel basis, rescaling the continuum model to the median of the observed continuum in each spaxel. 

The continuum subtracted datacube was finally used to fit the emission lines in each spaxel. The velocity and widths of the Gaussian components were bound to be the same for each emission line of the different species, while the intensities were left free to vary with the exception of the [NII]$\lambda$6548,84 and [OIII]$\lambda$4959,5007 doublets, were the intrinsic ratio between the two lines was used. We verified that this assumption, corresponding to assume that the different emission lines come from similar environments in the galaxy, applies in our data randomly inspecting the line profiles of the different emission lines in several spaxels (see also Fig.~\ref{bluebubble} for an extreme case).
Each fit was performed three times, with 1, 2 and 3 Gaussian components per each emission line, in order to reproduce peculiar line profiles where needed.  A selection based on the reduced $\chi^2$ obtained using each of the three different fits in each spaxel was used to select the spaxels where a multiple components fit was required to improve the fit. This choice allows us to use the more degenerate multiple components fits only where it is really needed to reproduce the observed spectral profiles: these conditions apply only in the central part of the galaxy and in some knots of the outflowing filaments where double peaked lines are detected, as discussed in the next Section.
 
\section{Gas and stellar dynamics} \label{dynamics}

\begin{figure*}
	\begin{center}
		\includegraphics[width=1\textwidth]{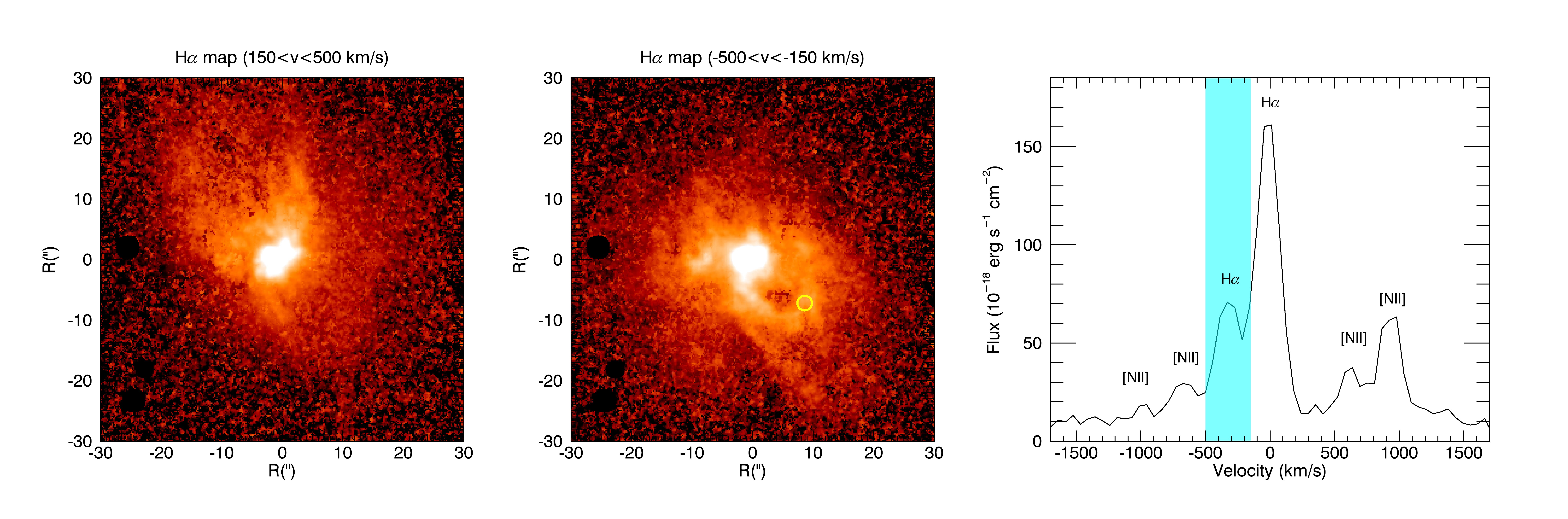} 
	\end{center}
	\caption{\textit{Left panel:} \ha\ channel map between $150<v<500\ km/s$, showing the redshifted expanding bubble to the NE. \textit{Central panel:} channel map between $-500<v<-180\ km/s$, showing the corresponding blueshifted bubble to the SW. The yellow circle shows the spaxels from which the spectrum of the [NII] and \ha\ region showed in the \textit{right panel} is extracted. The double peaked profile of \ha\ and [NII] is evident, with a shift between the two peaks of $310\ km/s$. The shaded cyan region shows the velocity range used for the channel map shown in the central panel. } 
	\label{bluebubble}
\end{figure*}

The velocity (first moment) and velocity dispersion (second moment) of the total emission line profiles in each pixel are used to study the gas dynamics in He~2-10. Fig.~\ref{Hadyn} shows the dynamical maps for the whole MUSE field of view: the most striking features are the high velocity and high velocity dispersion regions corresponding to the bubble structure NE (redshifted, Fig.~\ref{bluebubble}, left panel) and SW (blueshifted, Fig.~\ref{bluebubble}, central panel) of the center, evident in the \ha\ maps (see e.g. Fig.~\ref{3colors}). These structures are very extended, up to $\sim18''$ ($\sim720$ pc) projected from the central star forming regions, both to the NE and to the SW. The multiple bubbles form a complex structure (see Fig.~\ref{3colors}), that can be interpreted as due to multiple gas ejections from the central highly star forming region. 
While the bulk of the gas has velocity dispersions of $\sim50-60\ km/s$, corrected for instrument broadening, the bubbles show dispersions as high as $\gtrsim 180\ km/s$. The line profile is in fact broadened towards blue (SW) or red (NE) velocities, with the first moment of the lines clearly showing high velocity shifts, up to $\pm 130\ km/s$. The line profile in the bubbles region is particularly complex, showing double peaks with velocity differences as high as $\sim 310\ km/s$ (see Fig.~\ref{bluebubble}, right panel), consistent with the blueshift inferred from UV interstellar features detected by Johnson et al. \cite{johnson00}. In the following, we will assume that this velocity shift is representative of the average outflow velocity $v_{out}$, i.e. the de-projected velocity of the outflow, and regard the lower velocities observed as due to projection effects. Given that the escape velocity from He~2-10 is $v_{esc}\approx160\ km/s$ (Johnson et al. \citealp{johnson00}), it is clear that most of the gas detected in the bubbles has enough kinetic energy to escape the potential of the galaxy. Given the measured radius and velocity of the outflow, the dynamical time of the outflow is $t_d \sim R_{out}/v_{out} = 2.3\ Myr$.

In order to compute the total gas mass in the ionized outflow, we extracted a stellar continuum subtracted spectrum integrated in an aperture of 120 spaxels in radius ($24''$ or 960 pc), covering the full extent of the outflowing bubbles. We fit the \ha\ emission with a set of two Gaussian profiles, a broad ($FWHM_{b}=311\ km/s$; $F_{b}(H\alpha)=3.47\pm0.15\times10^{-12}\ erg\ s^{-1}\ cm^{-2}$) and a narrow component ($FWHM_{n}=148\ km/s$; $F_{n}(H\alpha)=1.93\pm0.07\times10^{-12}\ erg\ s^{-1}\ cm^{-2}$). Using the average derived extinction value on the outflow region from the Balmer decrement (see Sect.~\ref{avne}, $A_V\sim0.5$), we derived an extinction corrected luminosity for the broad component of $L_{b}(H\alpha)=1.77 \times 10^{40}\ erg\ s^{-1}$.  We note that this is a lower limit, as especially part of the redshifted component may be more absorbed by the intervening dust in the interposed galaxy. We assume that the broad component in the integrated spectrum is fully due to the outflowing bubble detected as kinematical features in the analysis above. As comparison, a stricter lower limit on the flux from the outflowing bubbles is given by a spatial analysis where we integrate the \ha\ emission only in the external regions where the high velocity gas is detected: in this case we find that the \ha\ flux is $\sim7\times10^{-13}\ erg\ s^{-1}\ cm^{-2}$, but missing all the emission from high velocity gas in the central region of the galaxy. Assuming the simplified outflow model by Genzel et al. \cite{genzel11}, we find that the mass in the ionized outflow is given by:
\begin{equation}
	M_{out}=3.2\times10^5\ \left(\frac{L_{b}(H\alpha)}{10^{40}\ erg/s}\right)\ \left(\frac{100\ cm^{-3}}{n_e}\right)\ M_{\sun} = 2.3\times10^5\ M_{\sun}
\end{equation}
Although this value is only taking into account the warm ionized gas, it is in broad agreement with the lower limit derived from UV absorption spectroscopy by Johnson et al. (\citealp{johnson00}, $M_{out}\approx10^5-10^6$) for the cold component. Assuming a biconical outflow distribution with velocity $v_{out}=310\ km/s$ out to a radius $R_{out}=720$ pc for the ionized wind, uniformly filled with outflowing clouds, the mass outflow rate is given by (see Cresci et al. \citealp{cresci15a}):
\begin{equation}
	\dot M_{out} \approx \mean{\rho_{out}}_V \cdot \Omega\ R^2_{out} \cdot v_{out} = 3\ v_{out}\ \frac{M_{out}}{R_{out}} = 0.30\ M_{\sun}\ yr^{-1}
\end{equation}
where $\mean{\rho_{out}}_V$ is the volume-average density of the gas and $\Omega$ the opening angle of the (bi-)cone. It is interesting to compare this outflow value with the SFR in the galaxy. The ratio between the two, i.e. the mass loading factor $\eta$, for the outflow in He~2-10 is therefore:
\begin{equation}
	\eta=\dot M_{out} / SFR = 0.39
\end{equation}
using the total $SFR=0.76\ M_{\sun}\ yr^{-1}$ from \ha\ emission in our MUSE data (see Sect.~\ref{gasexc}). This value for the mass loading factor is in the range derived by Arribas et al. \cite{arribas14} for local LIRGs ($\mean{\eta}=0.3$) and ULIRGs ($\mean{\eta}=0.5$), confirming the extreme nature of the starburst in this galaxy. 

The total kinetic energy in the outflow is $E_{out}(kin)\sim2.2\times10^{53}\ erg$, an order of magnitude lower than the energy of $3.5\times10^{54}\ erg$ that the $\sim3750$ Supernova Remnants (SNR), estimated by M{\'e}ndez et al. \cite{mendez99} from radio observations, can inject in the interstellar medium. An additional, comparable energy contribution is expected from stellar winds by massive stars (M{\'e}ndez et al. \citealp{mendez99}). The difference is probably due to radiative losses during the expansion of the bubble, as well as to the fact that our outflowing mass estimate is probably a lower limit. In any case, the starburst present in the central region of the galaxy is able to provide all the energy required to sustain the observed outflow.
\begin{figure*} 
	\begin{center}
		\includegraphics[width=0.7\textwidth]{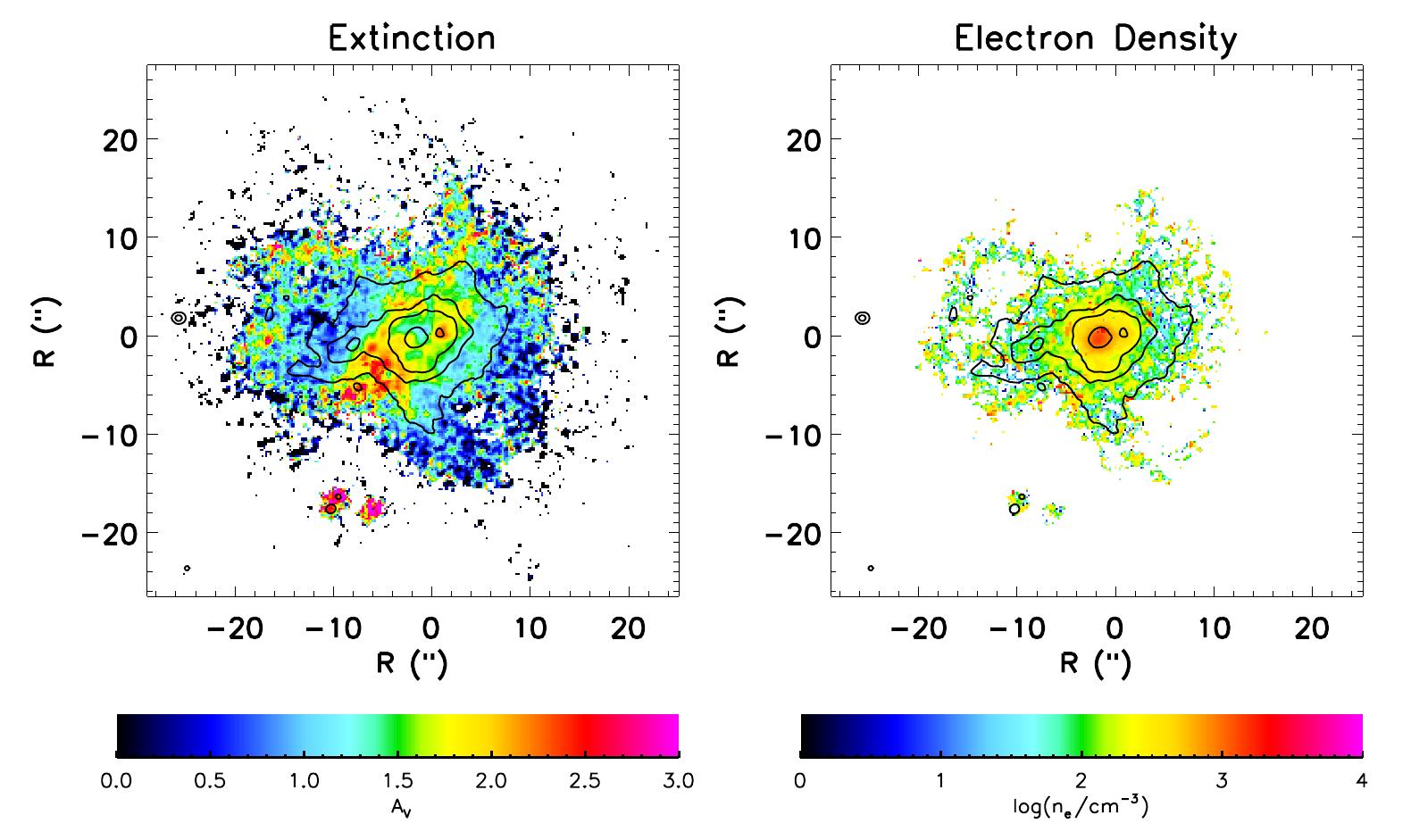}  
		\caption{Extinction map as derived from the \ha/\hb\ line ratio (\textit{left panel}) and electron density from \siia/\siib\ (\textit{right panel}), with \ha\ contours overplotted. Both maps are displaying the spaxels with $S/N>3$ in all the emission lines involved.}
		\label{dustdensity}
	\end{center}
	
\end{figure*}

In addition to the velocity structures due to the expanding bubbles, there is a velocity gradient in the SE (blueshift) - NW (redshift) direction, in the regions with lower velocity dispersion, consistent both in amplitude and orientation with the velocity field of the HI disk reported by Kobulnicky et al. \cite{kobulnicky95}. This is therefore the first detection of a large scale rotating warm ionized gas body in He~2-10.  

The star kinematics from stellar absorption features is instead different from what we derived for the ionized gas. Although the S/N on the continuum is much lower than in the line emission, the stellar velocity and velocity dispersion mapped in the Voronoi bins described in Sect.~\ref{lines} are mostly flat, with $\sigma\sim45\ km/s$ and no detectable velocity gradient or obvious signatures of past merging events, with velocity differences $\lesssim10\ km/s$. This is in agreement with the smaller field of view but higher S/N and spectral resolution IFU observations of Marquart et al. \cite{marquart07}, as well as the near-IR data of Nguyen et al.\cite{nguyen14}, confirming that the stars and the gas in this galaxy are at least partly dynamically decoupled in this galaxy.

\section{Gas properties} \label{gasmet}

The excitation, physical conditions, dust and metal content of the interstellar gas in He~2-10 can be explored using selected ratios between the measured emission lines. Thanks to our IFU observations, we are able to spatially map the line emission (see e.g. Fig.~\ref{3colors}) and line ratio diagnostics across the MUSE field of view. 

\subsection{Dust extinction and electron density} \label{avne}

First of all we use the Balmer decrement \ha/\hb\ to derive the dust extinction map, assuming a Calzetti et al. \cite{calzetti00} attenuation law and a fixed temperature of $10^4\ K$: the resulting extinction map is shown in Fig.~\ref{dustdensity} (left panel), for the spaxels where the \hb\ line was detected with $S/N > 3$. The dust attenuation appear to be highest in the two star forming clumps detached $\sim18''$ to the SW of the main galaxy and in the Eastern region where the line and continuum emission is less prominent. These locations correspond to the position of the CO gas (Kobulnicky et al. \citealp{kobulnicky95}). Moreover, the extinction is high on one of the central star forming regions ($A_V=2.3$, knot 1+2 in the radio notation by Kobulnicky \& Johnson \citealp{kobulnicky99}). It is interesting to compare this extinction map with the one obtained by near-IR IFU data for the central region of the galaxy by Cresci et al. \cite{cresci10} from the Br12/Br$\gamma$ line ratio, where the extinction towards the two brightest star forming region at the center of the galaxy was $A_V\sim7-8$. The difference is probably due to the fact that IR observations are capable to probe deeper in the highly embedded star forming clusters. 

The electron density was estimated using the \siia/\siib\ ratio (e.g. Osterbrock \& Ferland \citealp{osterbrock06}). We compute the line ratio in each spaxel where the [SII] lines are detected with $S/N>3$, and convert it to an electron density using the \textrm{IRAF} task \texttt{temden}, assuming a temperature of $10^4\ K$ as before. The resulting electron density map is shown in Fig.~\ref{dustdensity} (right panel). In this case, the highest density is obtained in the Western central star forming region (knot 4) where densities of $n_e=1500\ cm^{-3}$ are reached. The density distribution is instead flat in the rest of the galaxy, with densities $\sim 100\ cm^{-3}$.

This evidence confirms that the central regions of He~2-10 host dense, dust embedded, young and highly star forming star clusters, a common feature in starburst galaxies (see e.g. Vanzi \& Sauvage \citealp{vanzi06}) and possibly in galaxies in general (e.g. F{\"o}rster Schreiber et al. \citealp{fs11}).    

\begin{figure*}
	\begin{center}
		\includegraphics[width=0.6\textwidth]{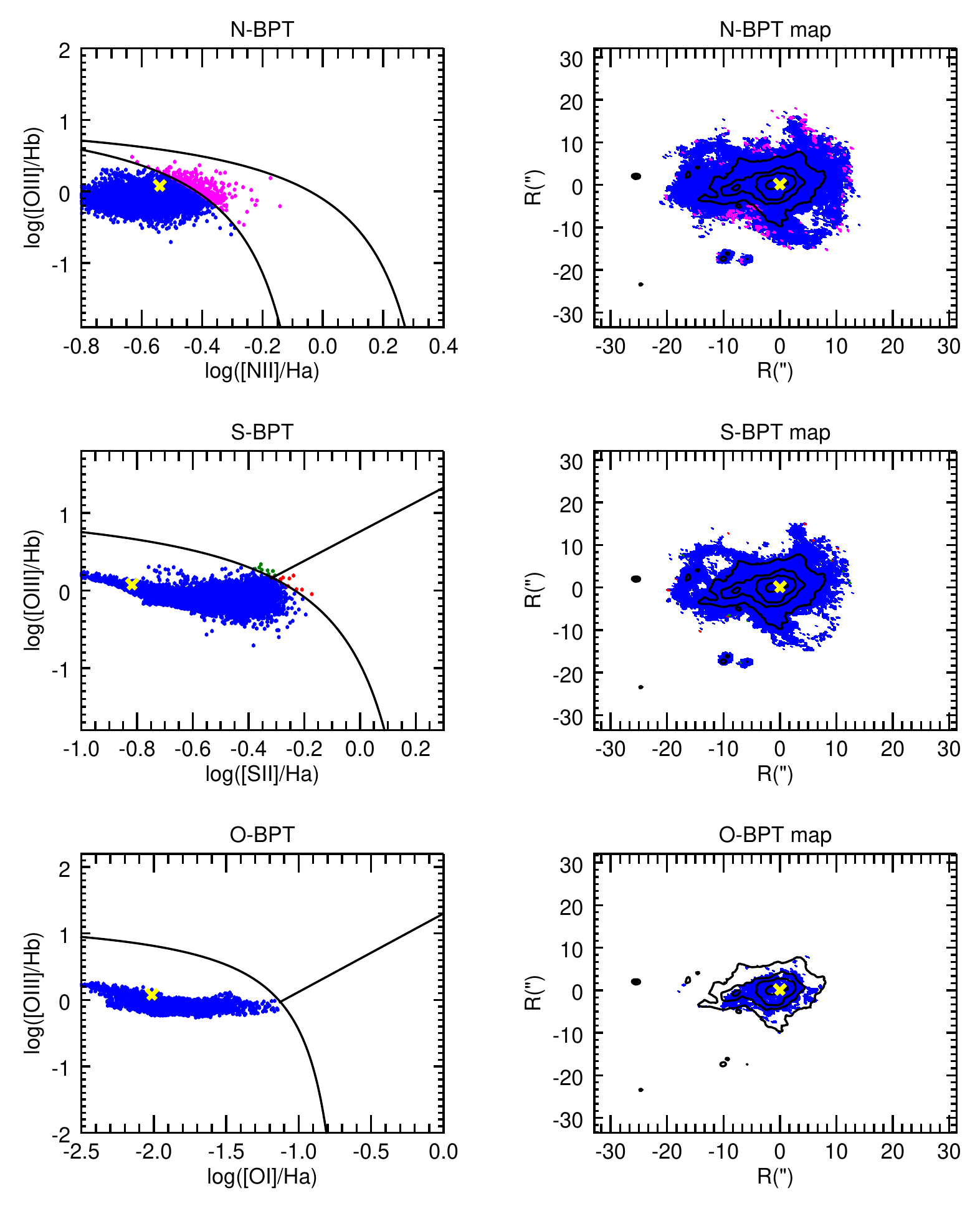}
	\end{center}
	\caption{Resolved BPT diagrams for He~2-10. The N-BPT (\niib/\ha\ vs \oiiib/\hb\, upper panels), S-BPT (\siiab/\ha\ vs \oiiib/\hb\, central panels) and O-BPT (\oi/\ha\ vs \oiiib/\hb\, lower panels) diagrams for each spaxel with $S/N > 3$ in each line are shown on the left. The location of line ratios at the location of the nuclear radio source classified as accreting BH by Reines et al. \cite{reines11} are shown by a yellow cross. A map marking each spaxel with the color corresponding to the dominant excitation at its location is shown on the right, again for spaxels with $S/N > 3$. The star forming regions are marked in blue, the Seyfert-type ionization is shown in green, LINER/shock dominated regions are shown in red and intermediate regions are shown in magenta in the N-BPT. The contours of \ha\ line emission are overplotted in black, and the location of the radio source is marked with a yellow cross. All the line emission is dominated by star formation like ionization in He~2-10, including at the location of the compact radio source.} 
	\label{bptall}
\end{figure*}

\subsection{Gas excitation} \label{gasexc}

We investigate the dominant ionization source for the line emitting gas in each MUSE spaxel using the so called BPT diagrams (Baldwin et al. \cite{baldwin81}). Along with the classical diagnostic using the \oiiib/\hb\ versus \niib/\ha\ line ratios (N-BPT in the following), we also explore the alternative versions using \siiab\ (S-BPT) or \oi\ instead of \niib\ (O-BPT, see e.g. Kewley et al. \citealp{kewley06}; Lamareille \citealp{lamareille10}). In all these diagrams, galaxies or spatially resolved regions of galaxies dominated by star formation, AGN (Seyfert-type), Low Ionization Emission line Regions (LIERs, Belfiore et al. \citealp{belfiore16}), or shocks populate different regions. The three BPT diagrams for the spaxels in He~2-10 were the $S/N>3$ for all the lines involved are shown in Fig.~\ref{bptall} (left panels). The dominant source of ionization is marked with different colors in each plot, blue for star formation, magenta for intermediate regions in the N-BPT diagram, green for AGN like ionizing spectra, red for LIER/shocks. The position of the different source of ionization in the map of He~2-10 is reported as well in Fig.~\ref{bptall} (right panels), where each spaxel is plotted with the color corresponding to its dominating ionization source. Clearly, all the line emitting gas in the galaxy is dominated by ionization from young stars, as different sources are limited to few noisier spaxels at the edges of the galaxy.  

Given that all the line emitting gas is ionized by young stars, we use an integrated stellar continuum subtracted spectrum, extracted in an aperture of 120 spaxels in radius ($24''$ or 960 pc) to compute the total SFR in the galaxy. We derive a total $L(H\alpha)=1.48 \times 10^{41}\ erg\ s^{-1}$, after correcting for the dust extinction using the Balmer decrement ($A_{H\alpha}=1.32$). This value for the \ha\ luminosity converts into a $SFR=0.76\ M_{\sun}\ yr^{-1}$, using the calibration by Kennicutt \& Evans \cite{kennicutt12}.

\begin{figure*}
	\begin{center}
		\includegraphics[width=0.7\textwidth]{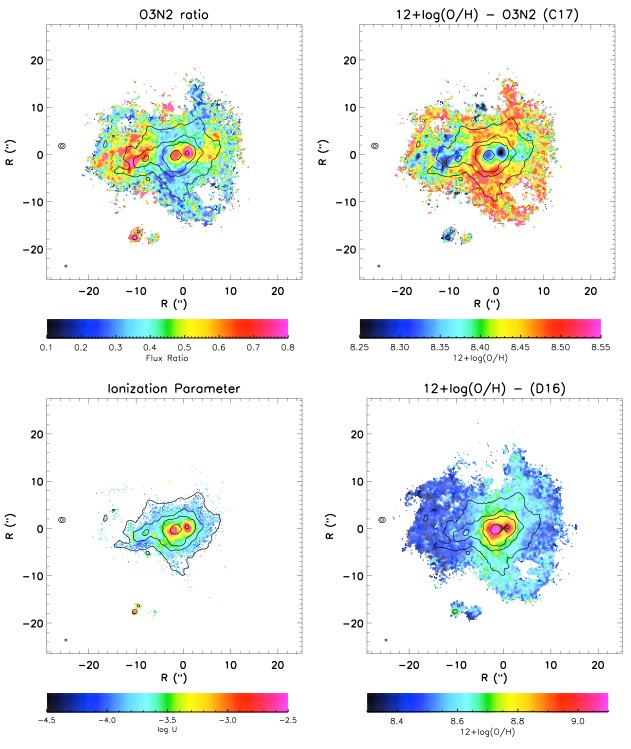} 
	\end{center}
	\caption{Spatial dependence of the $O3N2$ line ratio (\oiiib/\hb$\cdot$\ha/\niib, \textit{upper left panel}), derived metallicity (from $O3N2$, following Curti et al. \cite{curti17} calibrations, \textit{upper right panel}), ionization parameter $U$ from the \siii/\siiab\ line ratio (\textit{lower left panel}) and the metallicity map using the diagnostic by Dopita et al. \cite{dopita16} (\textit{lower right panel}). The central regions appear to have higher O3N2 and therefore lower metallicity according to single line ratio diagnostics such as $O3N2$, but the ionization parameter is one order of magnitude higher than the outer regions of the galaxy. The diagnostic by Dopita et al. \cite{dopita16} actually predict higher metallicity in those regions, and a much larger metallicity gradient. All maps are displaying the spaxels with $S/N>3$ in all the emission lines involved.}
	\label{ionization}
\end{figure*}

\subsection{Metallicity and ionization} \label{metion}

As according to diagnostic diagrams discussed in the previous section the gas ionization is dominated by young stars across all the galaxy, it is possible to use selected line ratios of the most intense lines to derive the chemical enrichment of the interstellar gas (the so-called ``strong line methods'', see e.g. Curti et al. \citealp{curti17} and references therein). The integrated metal abundance of He~2-10 was already derived by Kobulnicky et al. \cite{kobulnicky99b} using this method, yielding a super solar abundance of $12+log(O/H)=8.93$. Such a high value 
is not unexpected given the relative higher mass of He~2-10 and the well known relation between mass and metallicity: as an example, assuming the mass metallicity relation of Tremonti et al. \cite{tremonti04} and a stellar mass $M_*=3.7\times 10^9\ M_{\sun}$, we expect $12+log(O/H)\sim8.82$, that is increased to $8.95$ using the higher mass $M_*=1\times 10^{10}\ M_{\sun}$ by Nguyen et al. \cite{nguyen14}. If we also consider the SFR as third parameter in the relation,  Mannucci et al. \cite{mannucci10} would predict $12+log(O/H)\sim8.70$, somehow lower than measured. However, Esteban et al. \cite{esteban14} recently obtained a new measure of the oxygen metallicity integrated on the central $8''\times3''$ of $12+log(O/H)=8.55\pm0.02$, using faint pure recombination lines. Such difference between the different methods to measure metallicity and different calibrations is a well known effect (see e.g. the discussion in Kewley \& Ellison \citealp{kewley08}), and mostly due to a mismatch between direct methods, based on electron temperature $T_e$ or recombination lines, and strong line calibrations based on photoionization models, that are offset by $\sim0.3-0.5$ dex at high metallicity. 
\begin{figure*}
	\begin{center}
		\includegraphics[width=0.8\textwidth]{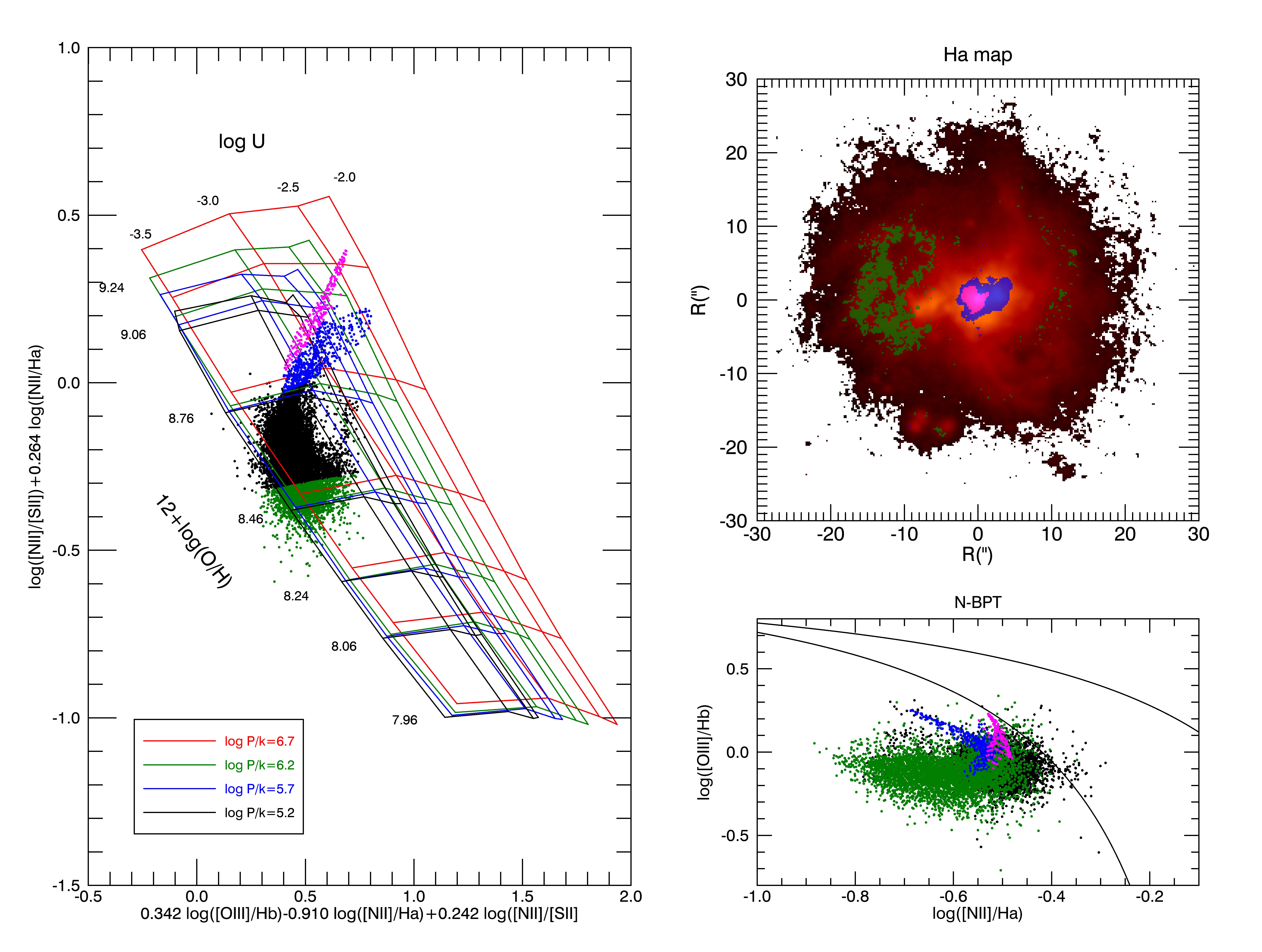}
	\end{center}
	\caption{The spatially resolved emission line diagnostic by Dopita et al. \cite{dopita16} for spaxels in He~2-10 with $S/N>3$ in all the emission lines involved. The diagnostic diagram is shown in the left panel, where each point correspond to a MUSE spaxel. The metallicity scale (increasing towards the upper part of the plot) and ionization parameter scale (increasing towards the right part) are shown on the different grids, corresponding to different values of the ISM pressure. The spaxels whose metallicity in the Dopita et al. \cite{dopita16} scale is lower than $12+log(O/H)=8.5$ are marked in green, while the two horn like structures at high metallicity and ionization parameter are shown in magenta and blue. These regions are marked with the same color in the upper right panel, overplotted on the \ha\ map of the galaxy: the high metallicity and ionization parameter regions correspond to the highly star forming clusters, while the low metallicity region of the galaxy is mostly in the Western part. The lower right panel shows the position of the regions selected on the diagnostic diagram on the N-BPT.} 
	 \label{dopita}
\end{figure*}
This mismatch is now being solved by the advent of new ``strong line'' diagnostics, fully empirical calibrated using direct methods up to high metallicities (see e.g. Brown et al. \citealp{brown16}, Curti et al. \citealp{curti17}). The new metallicity value by Esteban et al. \cite{esteban14} is again comparable with what is expected by a computation of the Mass Metallicity Relation obtained with such new diagnostics, as Andrews \& Martini \cite{andrews13} would predict $12+log(O/H)\sim8.65$ using only the stellar mass and $12+log(O/H)\sim8.50$ taking into account the SFR as well.

Here we attempt to spatially map the metallicity in He~2-10 using this new calibrations and our IFU data. In Fig.~\ref{ionization} (first panel) we show the spatial variation of one of these diagnostic ratios, O3N2=\oiiib/\hb$\cdot$\ha/\niib, and the corresponding metallicity map, using the calibrations by Curti et al. \cite{curti17} (second panel). It can be seen how the global average metallicity in the disk is compatible with the value derived by Esteban et al. \cite{esteban14} with recombination lines, while the Eastern region of the galaxy and the central highly star forming regions show lower metallicity values ($12+log(O/H)=8.2-8.3$). Highly star forming regions with lower gas metallicity than the surrounding galaxy have been interpreted as signatures of pristine, low metallicity gas accretion, that dilutes the metal content of the ISM and boosts star formation, both locally (e.g. S\'anchez Almeida et al. \citealp{SA14}) and at high-z (Cresci et al. \citealp{cresci10b}). 

However, in the case of He~2-10 the abundance gradient amplitudes is of the same order of magnitude of the intrinsic scatter in the calibrations (e.g. $0.2$ dex for O3N2, probably due to variation in the ionization in different sources). Moreover, it varies depending on the diagnostic ratio used: as an example, the gradient using \niib/\ha\ is $\Delta Z\sim0.15$ dex, using O32 is $\Delta Z\sim0.25$ dex, while using a simultaneous fit with a combination of the above diagnostics and of \oiiib/\hb\ is $\Delta Z\sim0.12$ dex, although the different indicators are supposed to be cross-calibrated. Moreover, we use the line ratio [SIII]/[SII] to derive a map of the ionization parameter $U$, using the calibrations of Kewley \& Dopita \cite{kewley02} (Fig.~\ref{ionization}, third panel). The ionization parameter is the ratio between the number density of photons at the Lyman edge and the number density of Hydrogen, $U=Q_{H^0}/(4\pi R_s^2\ c\ n_H)$, where $Q_{H^0}$ is the flux of ionizing photons produced by the exciting stars above the Lyman limit, $n_H$ the number density of Hydrogen atoms, $c$ the speed of light and $R_s$ the Str\"omgren radius of the nebula, and it is a measure of the intensity of the radiation field. It can be seen from Fig.~\ref{ionization} (third panel) that this parameter is one order of magnitude higher in the two nuclear star forming region, further suggesting that the line ratio variation is due to ionization effects in extreme environments such as these embedded, highly 
star forming regions and not to a metallicity variation. 

To further explore this, we map the MUSE spaxels on the new diagnostic diagram by Dopita et al. \cite{dopita16}, which makes use of \oiiib, \hb, \niib, \siiab\ and \ha\ lines to constrain both metallicity and $U$ using photoionization models. The result is shown in Fig.~\ref{dopita} (left panel): the different grids shows the variation of metallicity and $logU$ for different values of ISM pressure. The spaxels corresponding to the central star forming regions define two clear sequences towards high metallicity and high $U$, plotted in magenta and blue. The location of these spaxels are plotted in the upper right panel superimposed to the \ha\ map of He~2-10. The spaxels with metallicity $12+log(O/H)<8.55$ are marked in green, both in the left and in the right panel. This confirms that the Eastern part of the galaxy has lower metallicity than the rest by $\sim0.2\ dex$, possibily in agreement with the merger interpretation of the origin of He~2-10, while the different line ratios in the central star forming regions are due to higher ionization parameter (see also Fig.~\ref{ionization}, fourth panel). In the Dopita et al. \cite{dopita16} diagram the two star forming regions actually show very high supersolar metallicities, probably due to efficient metal enrichment in those extreme environments. As shown in Fig.~\ref{dopita}, lower right panel, these regions form a definite structure in the BPT diagram as well. This result suggests caution in the interpretation of a single line ratio as a variation in metallicity of the ISM, and confirms the importance of a large wavelength range to exploit multiple physical diagnostics.

\begin{figure*}
	\begin{center}
		\includegraphics[width=1\textwidth]{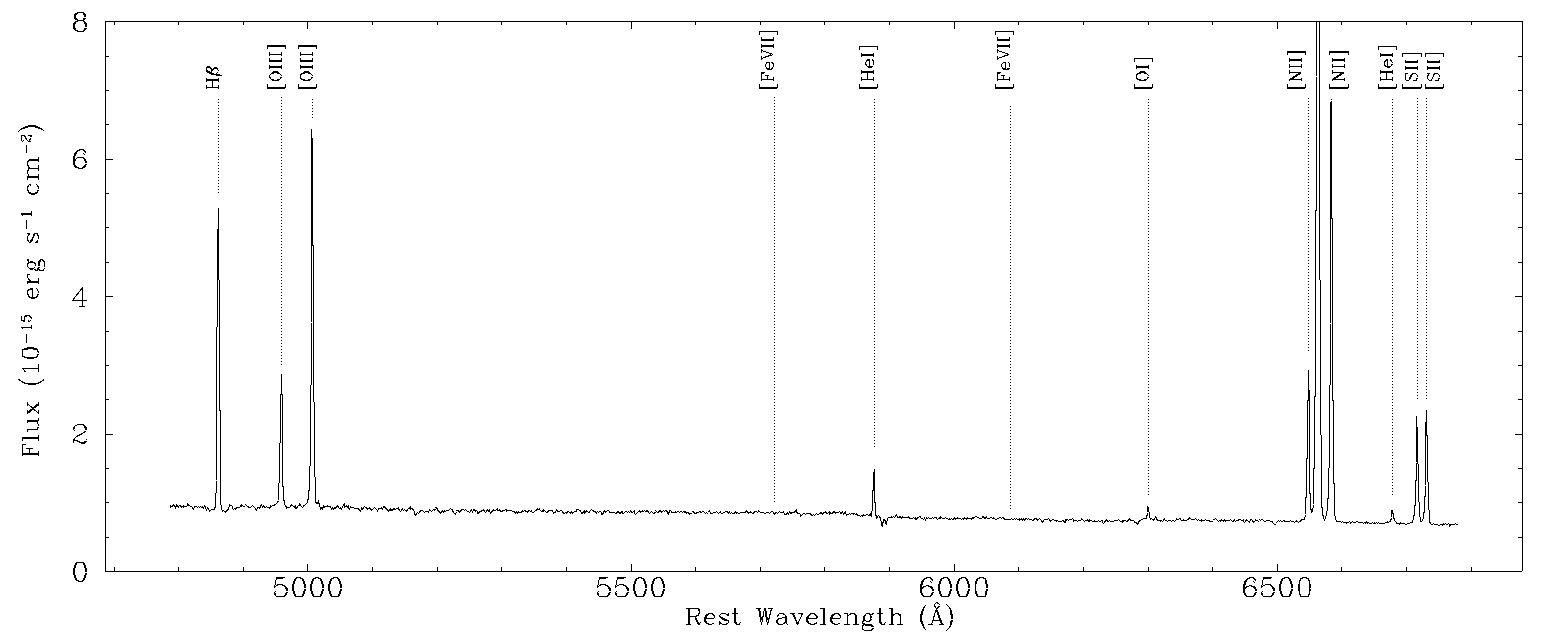} 
	\end{center}
	\caption{Spectra integrated over an aperture of 2 pixels radius ($0.4"$) around the location of the compact radio source and putative black hole in HE~2-10. The main emission line detected are labelled, as well as the expected position of higher ionization emission lines typical of AGN spectra such as [FeVII], that are however non detected in He~2-10.}
	\label{bhspectrum}
\end{figure*}

\section{No coherent evidence for a SMBH in He~2-10: a Supernova Remnant origin for the compact radio emission?} \label{BH}

As discussed in the introduction, Reines et al. \cite{reines11} identified as an accreting SMBH the compact radio source detected by Johnson \& Kobulnicky (\citealp{johnson03}, their knot 3 located between the two strongest star forming regions), based on the combination of radio and X-ray data. In particular, they used the ratio between the radio to X-ray luminosity, $R_X=\nu L_{\nu} (5\ GHz)/L_X(2-10\ keV)$ to discriminate between different scenarios: their data provided $log R_X\sim-3.6$ for the radio source 3, that is in the range expected for low luminosity AGNs ($log R_X\simeq-2.8\div-3.8$, Ho et al. \citealp{ho08}). A stellar-mass black hole X-ray binary system was excluded, as this kind of sources are too weak in the radio ($log R_X<-5.3$) to explain the observed ratio. Conversely, Supernova Remnants (SNR) were excluded as they are relatively weak X-ray sources, with $log R_X\gtrsim-2.7$. 

We therefore extracted a spectrum integrated over an aperture of 2 pixels radius ($0.4"$) around the location of the compact radio source 3 and putative black hole to check for any sign of AGN ionization in the MUSE data. The spectrum is shown in Fig.~\ref{bhspectrum}. The position of the line ratios extracted from this spectrum on the BPT diagrams is shown in Fig.~\ref{bptall} as a yellow cross, showing that the ionizing radiation is completely dominated by young stars. Moreover, despite the very high S/N obtained, no high ionization lines typical of AGNs such as [FeVII]$\lambda$5721 and [FeVII]$\lambda$6087 are detected (e.g. Vanden Berk et al. \citealp{vandenberk01}). We derive an upper limit of $7\cdot10^{-3}$ for the [FeVII]$\lambda$6087/\hb\ ratio, while it is typically $\sim0.1$ in Seyfert galaxies (see e.g. Netzer \citealp{netzer90}). We therefore conclude that even if a SMBH is present in He~2-10, it is not contributing to the gas ionization.

Deeper (200 ks) Chandra observations recently obtained by Reines et al. \cite{reines16} have shown, on the basis of an accurate spectral, spatial and temporal analysis, that the hard X-ray radiation previously attributed to the compact radio source 3 in the shallower archive data (20 ks) is instead due to a varying source located in one of the star forming region (knot 4), compatible with a massive X-ray binary. They also reported the discovery of a faint source coincident with the radio source 3, with a tentative detection of a $\sim9$ hours periodicity which can in principle be ascribed to instabilities in the accretion disc flow, although the light curve can be adequately reproduced by a simple constant value. Its X-ray spectrum is very steep, and the $\sim180$ X-ray counts are fitted with an unobscured power law with $\Gamma$=2.9. However, as noted also by Reines et al. \cite{reines16}, the low counting statistics and the high local background due to the extended emission does not allow an unique characterisation of the X-ray emission, given that also a thermal plasma model with kT$\sim1.1$ keV can also well reproduce the data.

The luminosity associated with the radio source 3 over the full (0.3-10 keV) X-ray range, assuming the $\Gamma$=2.9 fit, is  $\sim10^{38}\ erg\ s^{-1}$. This translates into an hard (2-10 keV) X-ray luminosity of $L_{2-10 keV}\sim1.4\times10^{37}\ erg\ s^{-1}$. 
We reanalyzed the spectrum of the radio source 3 in the longest Chandra observation available (160 ks), extracted from a 0.5$\arcsec$ radius circle around the radio centroid. We retrieve a consistent hard X-ray luminosity when using the same power-law model adopted by Reines et al. \cite{reines16},  but with residuals at E$>5$ keV at $\sim 3\sigma$. We therefore conservatively added an additional power law component to the fit, in the hypothesis that we see emission from an (obscured) AGN. This returns an hard X-ray luminosity of $1.0\times10^{38}$ erg s$^{-1}$.
However, a contamination from the variable high mass X-ray Binary north of the nuclear source is still present, as all the detected photons with E$>5$ keV are located in a $0.5''$ radius from the contaminating source. For these reasons, the $L_X$ derived from this two components fit should be considered as a very conservative upper limit to the maximum 2-10 keV luminosity allowed for the radio source. 


The upper limit on the hard X-ray luminosity adding a second power law to the fit corresponds to a radio to X-ray ratio $R_X>-2.4$, assuming the revised $\nu L_{\nu} (5\ GHz)=4.1\times10^{35}\ erg\ s^{-1}$, an order of magnitude higher than what is measured in local low luminosity AGNs, and even a factor of 2.5 higher than transitional objects between HII galaxies and LINERs (Ho et al. \citealp{ho08}). A much higher value of $R_X=-1.6$, more than an order of magnitude higher than any kind of local low luminosity AGN, is instead obtained from the single power-law fit X-ray luminosity. 
Both these new values for $R_X$ are actually now compatible with the range observed in SNRs ($R_X\gtrsim-2.7$, Reines et al. \citealp{reines11}, Vink \citealp{vink12}). This scenario is further supported by the detection of high [FeII]/Br$\gamma$ at the location of the compact radio source by Cresci et al. \cite{cresci10}, consistent with thermal excitation in SNR shocks. In this respect we note that a high [FeII] emission also corresponds to the location of a second compact radio source in He~2-10 detected by Reines \& Deller \cite{reines12} and therein classified as a SNR. Moreover, the radio spectral index $\alpha\sim-0.5$ (Reines \& Deller \citealp{reines12}) is the typical spectral index measured in SNR (e.g. Reynolds et al. \citealp{reynolds12}), and its radio luminosity is high but still compatible with a young SNR, as well as its compact size (see e.g. Fenech et al. \citealp{fenech10}).

On the other hand, if we assume that both the radio and hard X-ray luminosity is due to a SMBH, the fundamental plane of BH activity (Merloni et al. \citealp{merloni03}) would predict a BH mass of $log(M_{BH}/M_{\sun})=7.65\pm0.62$ using the single component X-ray spectrum fit. This BH mass would be incompatible with the stellar dynamical measurements obtained by Nguyen et al. \cite{nguyen14} with near-IR IFU observations at Adaptive Optics assisted spatial resolution ($0.15"$, comparable with the sphere of influence of a $10^6\ M_{\sun}$ BH of 4 pc, given the local stellar $\sigma=45\ km\ s^{-1}$). They placed a firm upper limit of $log(M_{BH}/M_{\sun})<7$ at $3\ \sigma$ level based on their data, finding no dispersion peak or enhanced rotation around the compact radio source, where actually the dispersion is lower than the surrounding area. 
Such a massive BH is also excluded by measurements of the dynamical map in the central 70 pc by CO observations of Kobulnicky et al. \cite{kobulnicky95} ($3-48\cdot10^6\ M_{\sun}$). We note that if we assume the maximum 2-10 keV luminosity allowed from the fit with an obscured AGN component ($L_X<1\times10^{38}$ erg s$^{-1}$), the Merloni et al. (2003) relation gives a lower limit to the SMBH powering the AGN, $log(M_{BH}/M_{\sun})>7$, that is still ruled out by the dynamical measurements.

Summarizing, the gas excitation in the MUSE data do not show any evidences of an active BH in He~2-10. This is further supported by the analysis of [FeII]/Br$\gamma$ ratio suggesting thermal excitation in SNR shocks. Moreover, the new Chandra observations that revised the hard X-ray flux from the putative location of the BH provide a flux $\sim 200$ times smaller than the estimate reported in Reines et al. \cite{reines11} based on the first 20 ks of Chandra data. The corresponding luminosity is now compatible with a SNR origin for the non thermal emission, and so it is the Radio to X-ray loudness ratio ($>-2.4$), much higher than that observed in local low luminosity AGNs. In case the X-ray and radio emission are still attributed to an accreting BH, the resulting  $log(M_{BH}/M_{\sun})=7.65$ would be too large to fit the dynamical information available, as well as the known correlations between BH mass and bulge mass (e.g. Kormendy \& Ho \citealp{kormendy13}), given the dynamical mass of $\sim7\cdot10^9\ M_{\sun}$ (Kobulnicky et al. \citealp{kobulnicky95}). Although a smaller BH would be still allowed by the large scatter in the relations, the resulting Eddington ratio for this putative SMBH  with $log(M_{BH}/M_{\sun})\sim7$ would be $L_{bol}/L_{Edd}\lesssim1\times10^{-6}$, at least two order of magnitudes lower than the value adopted for actively accreting BHs (see e.g. Merloni \& Heinz \citealp{merloni08}). Therefore, an actively accreting SMBH origin for the radio source 3 is excluded, and a young SNR seems easier to fit the gathered evidences. The only feature that cannot be simply ascribed to a SNR is the periodicity observed in the X-rays which, however, is reported as tentative and possibly due to random fluctuations (Reines et al. \citealp{reines16}). Further high resolution multi-wavelength observations are needed to definitely assess the nature of this interesting system.

\section{Conclusions} \label{conclusions}

We have presented MUSE integral field observations of He~2-10, a prototypical nearby HII galaxy. The MUSE data allowed a detailed study of the dynamics and physical conditions of the warm ionized gas in the galaxy. Our main results can be summarized as follows:

\begin{itemize}
\item[-] The gas dynamic is characterized by a complex and extended (up to 720 pc projected from the central star forming region) high velocity expanding bubbles system, with both blue and redshifted gas velocities $v_{max}> 500\ km/s$, higher than the galaxy escape velocity. We derive a mass outflow rate $\dot M_{out}\sim 0.30\ M_{\sun}\ yr^{-1}$, corresponding to mass loading factor $\eta\sim0.4$, in range with similar measurements in local LIRGs. Such a massive outflow has a total kinetic energy that is sustainable by the stellar winds and Supernova Remnants expected in the galaxy, with no additional energy source required.
\item[-] The lower velocity dispersion regions ($\sigma\sim55\ km/s$), where the outflowing gas is not the dominant component, show a velocity gradient of $\pm 35\ km/s$ consistent with the HI disk detected in He~2-10, thus tracing a rotating gaseous disk. 
The stellar kinematics is instead remarkably flat, suggesting a decouple between the bulk of the stellar population in the galaxy and the warm ionized gas.
\item[-] The central star forming regions where the most embedded star cluster reside show the highest values of electron density ($n_e=1500\ cm^{-3}$) and ionization parameter ($logU=-2.5$), confirming the extreme conditions in these environments. The dust extinction as derived by the Balmer decrement is high in the Western star forming knot ($A_V=2.3$) and in a region to the SE of the nucleus, corresponding to the location of an accreting CO cloud ($A_V\sim2.5$). Higher values of the dust extinction are obtained in the near-IR, probably because the optical diagnostic are just looking at the external shells of the embedded star forming regions. 
\item[-] Given the large variation of the ionization parameter across the galaxy, the use of a single line ratio to derive the metallicity of the galaxy is not recommended, as we show that the line ratio variation is due to a combination of metallicity and ionization variations. We use a diagram proposed by Dopita et al \cite{dopita16} to account simultaneously to the variation of these two parameters, finding a large metallicity gradient across the galaxy, with the central star forming regions having super solar metallicity, while the external part of the galaxy chemical abundances as low as $12+log(O/H)\sim8.3$.
\item[-] The ionization all across the galaxy is dominated by young stars, as described by the so called BPT diagrams, with no sign of AGN ionization. This seems in contrast to the claim of the presence of an accreting SMBH in He~2-10 (Reines et al. \citealp{reines11}, \citealp{reines16}), based on the combination of radio and X-ray data. However, we show that the X-ray radio-loudness parameters $R_X$ obtained with a revised estimate of the X-ray flux are much higher than expected from local low luminosity AGN but consistent with a SNR origin, as suggested also by the high [FeII]/Br$\gamma$ ratio observed at the location of the compact radio source. 
Therefore, the constraints given by the hard X-ray flux are not enough to confirm the presence of a SMBH in He~2-10, as the data can be explained with different kind of sources such as a young SNR. 
\end{itemize}

This work confirms the unique capabilities of large field integral field spectroscopy to explore the details and physical properties of the interstellar medium and star formation in nearby starburst galaxies. Upcoming similar observations of a larger sample of nearby HII and dwarf star forming galaxies will allow to shed light on the different mechanisms that regulates and trigger the star formation in these extreme environments.

\section*{Acknowledgments}

MUSE data were obtained from observations made with the ESO Telescopes at the Paranal Observatory. We are grateful to the ESO staff for their work and support. We also show our gratitude to E. Amato, R. Bandiera and N. Bucciantini for useful suggestions and discussion on Supernova Remnants, to A. Comastri on X-ray spectral fitting, and to A. Marconi and G. Venturi for sharing the python scripts used for the fitting. LV acknowledges support by the project CONICYT Anillo ACT-1417. MB acknowledges support from the FP7 Career Integration Grant ``eEASy'' (CIG 321913). GC, MB and GL acknowledge financial support from INAF under the contracts PRIN-INAF-2014 (``Windy Black Holes combing Galaxy evolution"). GC is also grateful to F. Chiesa for reminding the importance of never giving up.


\end{document}